\documentclass[pdflatex,sn-mathphys-num]{sn-jnl}

\usepackage{graphicx}%
\usepackage{multirow}%
\usepackage{amsmath,amssymb,amsfonts}%
\usepackage{amsthm}%
\usepackage{mathrsfs}%
\usepackage[title]{appendix}%
\usepackage{xcolor}%
\usepackage{textcomp}%
\usepackage{manyfoot}%
\usepackage{booktabs}%
\usepackage{algorithm}%
\usepackage{algorithmicx}%
\usepackage{algpseudocode}%
\usepackage{listings}%

\theoremstyle{thmstyleone}%
\theoremstyle{thmstyletwo}%

\theoremstyle{thmstylethree}%

\begin{document}

\title[Article Title]{Deep Learning–Based Estimation of Blood Glucose Levels from Multidirectional Scleral Blood Vessel Imaging}


\author[1,2,3]{\fnm{Muhammad Ahmed} \sur{Khan}}\email{khan@aierchina.com}
\equalcont{These authors contributed equally to this work and share first authorship.}

\author[1,2]{\fnm{Manqiang} \sur{Peng}}\email{peng762550@sina.com}
\equalcont{Muhammad Ahmed Khan and Manqiang Peng contributed equally to this work and share first authorship.}
\author*[1,2,3]{\fnm{Ding} \sur{Lin}}\email{dlinoph@163.com}

\author*[4,5]{\fnm{Saif Ur Rehman} \sur{Khan}}\email{mr.saifurrehman.khan@gmail.com}

\affil[1]{\orgname{Changsha Aier Eye Hospital}, 
\orgaddress{\street{South Furong Rd}, \city{Changsha}, \state{Hunan}, \country{China}}}

\affil[2]{\orgname{Aier Eye Hospital Group Co., Ltd}, 
\orgaddress{\street{South Furong Rd}, \city{Changsha}, \state{Hunan}, \country{China}}}

\affil[3]{\orgname{Central South University AIER Eye Institute of Ophthalmology}, 
\orgaddress{\city{Changsha}, \state{Hunan}, \country{China}}}

\affil[4]{\orgname{Central South University}, 
\orgaddress{\city{Changsha}, \state{Hunan}, \country{China}}}

\affil[5]{\orgname{German Research Center for Artificial Intelligence}, \orgaddress{\city{Kaiserslautern}, \postcode{67663}, \country{Germany}}}

\abstract{Regular monitoring of glycemic status is essential for diabetes management, yet conventional blood-based testing can be burdensome for frequent assessment. The sclera contains superficial microvasculature that may exhibit diabetes related alterations and is readily visible on the ocular surface. We propose \textbf{ScleraGluNet}, a multiview deep-learning framework for three-class metabolic status classification (normal, controlled diabetes, and high-glucose diabetes) and continuous fasting plasma glucose (FPG) estimation from multidirectional scleral vessel images. The dataset comprised 445 participants (150/140/155) and 2,225 anterior-segment images acquired from five gaze directions per participant. 
After vascular enhancement, features were extracted using parallel convolutional branches, refined with \textbf{Manta Ray Foraging Optimization (MRFO)}, and fused via transformer-based cross-view attention. Performance was evaluated using subject-wise five-fold cross-validation, with all images from each participant assigned to the same fold. 
ScleraGluNet achieved \textbf{93.8\%} overall accuracy, with one-vs-rest AUCs of \textbf{0.971}, \textbf{0.956}, and \textbf{0.982} for normal, controlled diabetes, and high-glucose diabetes, respectively. For FPG estimation, the model achieved \textbf{MAE = 6.42 mg/dL} and \textbf{RMSE = 7.91 mg/dL}, with strong correlation to laboratory measurements (\textbf{$r = 0.983$; $R^{2} = 0.966$}). 
Bland Altman analysis showed a mean bias of \textbf{+1.45 mg/dL} with 95\% limits of agreement from \textbf{$-8.33$} to \textbf{$+11.23$ mg/dL}. These results support multidirectional scleral vessel imaging with multiview learning as a promising noninvasive approach for glycemic assessment, warranting multicenter validation before clinical deployment.}




\maketitle

\section{Introduction}

Diabetes mellitus (DM) is becoming an increasingly important global menace and is one of the primary causes of avoidable ill health and death. As of 2021, the International Diabetes Federation estimated there to be 537 million adults living with diabetes, with projections of 643 million by 2030 and 783 million by 2045~\cite{dhatariya2025footulcer,hu2025globalburden,ogle2025t1d}. Persistent hyperglycemia is the engine that fuels acute metabolic decompensation and accelerates the detrimental function of the micro and macro vascular system, resulting in the various complications of diabetes including diabetic eye disease, kidney disease, diabetic nerve disease, heart disease, and stroke. Because of this, there is the need for the early identification of hyperglycemia with consequent monitoring to assist in case stratification and guide tailored management. The current gold standard in the clinical measurement of hyperglycemia (which includes FPG, Oral Glucose Tolerance Testing (OGTT), and Glycated Hemoglobin (HbA1c)) is accurate, but hinges on the requirement of blood sampling, clinical space, and are then inconvenient for use on a regular basis~\cite{thewjitcharoen2025ogtt,liang2025hba1c,khambule2025gestational}. The use of a finger prick to gather blood is still considered to be the standard of self-monitoring glucose levels but can lead to pain, and a risk of infection, and can lead to decreased compliance~\cite{helleputte2025cgm}. Continuous Glucose Monitoring (CGM) systems can be applied to reduce the number of fingerstick an individual must do and can help to increase the level of detail of the tracking; however, they still have a moderate degree of cost and require the person to undergo subcutaneous sensor insertion. Because of these limits, there is a great deal of interest in developing monitoring systems that can operate without a blood draw or blood sampling.

The ocular vasculature allows for non invasive metabolic evaluation to be conducted with ease because the microvascular structures and their functionalities can be visualized and measured accurately. It is known that chronic hyperglycemia can lead to remodeling and dysfunction of microvascular structures, which can damage their autoregulation capabilities and their ability to perfuse tissue. Among all the tissues that have been imaged, the retina is the most studied, and deep learning systems have demonstrated their capability to accurately detect and screen for diabetic retinopathy. Additionally, deep learning systems that go beyond just grading diabetic retinopathy have exhibited the ability to predict cardiometabolic status and estimate glycosylated hemoglobin through the analysis of retinal fundus images, which suggests that the patterns of retinal microvasculature might reflect other important factors related to metabolic systems.
Most significantly, the anterior segment (bulbar conjunctival/episcleral-scleral vasculature) provides other practical benefits for monitoring with a camera. Contrasted with the retina, visualizing and accessing scleral/conjunctival microvessels is feasible across large areas~\cite{zhang2025microvascular} and can be achieved with cheaper imaging acquisition systems, which is valuable for scalable screening and telemedicine. Prior investigations have documented diabetes-related changes within the bulbar conjunctival microvasculature, such as modified vessel morphology and vessel tortuosity~\cite{mone2025microvascular,tian2025nanozyme} and objective changes measured through red free imaging of the conjunctiva~\cite{nguyen2025multimodal}. Recent investigations using microcirculation quantitative techniques and optical coherence tomography angiography (OCTA) provide further evidence of the conjunctival and interstitial scleral vascular systems' diabetes-related microangiopathy and alterations in perfusion and density~\cite{liu2025conjunctival,chen2025conjunctival}. Taken together, the preceding studies adequately serve as the basis for the hypothesis that scleral/conjunctival microvasculature may be an indicator of the microvascular and systemic glycemic burden.

Due to concurrent developments in computer vision and biomedical signal processing, there are now ways to estimate physiological signals non-invasively and passively through the analysis of images and waveforms. While there have been multiple studies surrounding telehealth-enabled diabetes management systems, the analysis of consumer-grade devices with embedded photoplethysmography (PPG) to estimate glucose levels and their combination with thermal imaging have been the most prominent. That said, the field lacks models that are robust to motion and illumination, as most of these devices have weak physiological coupling and face domain shifts with variability in devices and environments. Scleral imaging, on the other hand, sheds light on these obstacles by giving direct and real-time visualization of the microvasculature, enhancing the interpretability and strength of the models when combined with advanced signal processing and representation learning.

Despite the enfranchisement of the ocular surface domain, there is still a general lack of edge regarding the vein-driven analytics processes. First, there is a user single-view acquisition frame phenomenon where this lacks a discernible step engine holding. This can paralyze the progression of any spatial and ramified vascular reconfiguration remodeling processes. The second phenomenon is the scleral mediated vascular neural net, and it displays a claustrophobic view. Scanning the tissue planes superior, inferior, nasal, and temporal has disconnected yet complementing neural net vessels with region-specific anomalies. To address the effective exploitation of this coherent interleaving of multi axes, a full view fused architecture is required. Additionally, ocular surface exposures converge precisely along the planes, and filament reconstructions, existing constellations, and design iterations amplify and refine the complex illumination and contrast variations across the ocular-surface images with unidirectional and variable channels. Finally, increasingly, MRFO has been the most effective and widely used engineering heuristic and, if used in engineering the multi-scalar vascular representations, can offer bioinspired efficiency of design.

To fill these gaps, we have developed \textbf{ScleraGluNet}, a multiview deep-learning model that estimates blood glucose levels and classifies metabolic states (normal, controlled diabetes, high glucose diabetic) from images of scleral vessels taken from five predefined gaze angles. This model contains: (i) a preprocessing step that enhances the vascular signals in the images; (ii) feature extraction via multibranch CNNs; (iii) MRFO feature refinement; and (iv) cross-view fusion with transformers to obtain global vascular signatures linked to glycemic status. Our model leverages the recent findings that external photographs of the eye can capture systemic biomarkers and indicators of glycemic control~\cite{adamu2025chaotic,spea2025cost}, and that augmenting the number of views can improve model robustness and generalization. Using a dataset of 2,225 images from 445 subjects, we report classification and regression results, along with ROC and agreement analyses and ablation studies to measure the impact of each architectural component.

\section{Materials and Methods}

\subsection{Population and Study Design}
This research was conducted as a single center observational study at Changsha Aier Eye Hospital. A total of 445 participants were consecutively enrolled from October 2025 to November 2025. The participants were divided into three groups based on their medical history and laboratory confirmed results.
\begin{itemize}

\item \textbf{Normal group (n = 150):} Participants in this group had no previous diagnosis or medical history of diabetes. Their FPG and HbA1c levels were within normal laboratory reference ranges.
\item \textbf{Controlled diabetes group (n = 140):} This group consisted of patients previously diagnosed with diabetes mellitus who were under medical supervision and demonstrated adequate glycemic control based on clinical assessment and laboratory findings.
\item \textbf{High glucose diabetic group (n = 155):} This group consisted of patients previously diagnosed with diabetes mellitus who showed poor glycemic control according to clinical assessment and laboratory results. Their diabetes related indices remained significantly elevated.
\end{itemize}

\begin{figure}[h]
    \centering
    \includegraphics[width=0.9\textwidth]{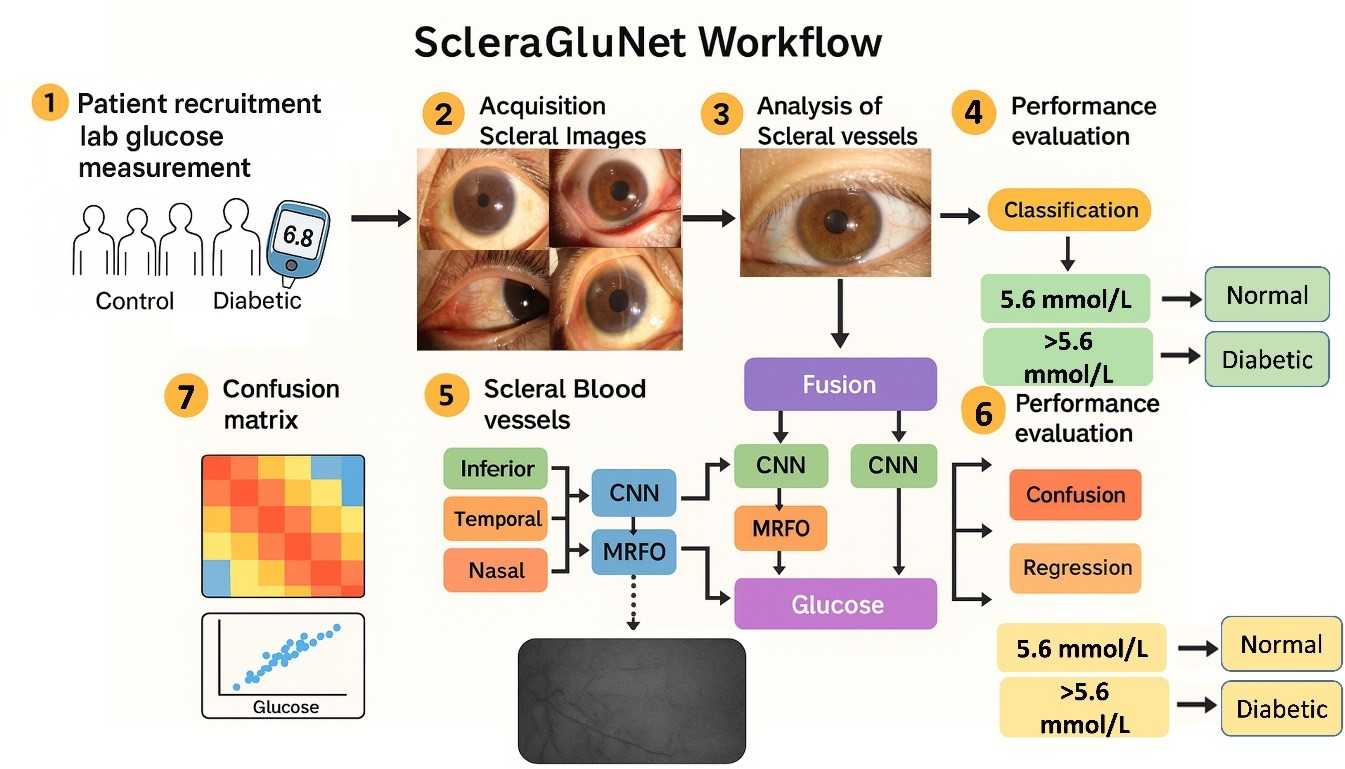}
    \caption{Overview of the study workflow}
    \label{fig1}
\end{figure}

On the day of imaging, each participant’s FPG and HbA1c values were recorded and used as reference standards for training and evaluating the model. Demographic information including age and sex was also collected for analysis. Participants were excluded if they had ocular surface diseases, active eye infections, recent ocular surgery or trauma, severe dry eye disease, or any other anterior segment conditions that could obscure the sclera. These exclusion criteria ensured accurate anterior segment imaging and reliable vascular assessments. The study followed the principles of the Declaration of Helsinki. Ethical approval was obtained from the Institutional Review Board of Changsha Aier Eye Hospital, and informed consent was obtained from all participants before inclusion in the study.
The overall experimental workflow from participant enrollment to model evaluation is illustrated in Fig.~\ref{fig1}. The experiment began with laboratory measurement of participants' FPG, which served as the ground truth reference. Subsequently, multi-directional scleral color photographs were captured from five gaze positions:

\begin{itemize}
\item Primary gaze (straight)
\item Superior gaze (up)
\item Inferior gaze (down)
\item Nasal gaze (left)
\item Temporal gaze (right)
\end{itemize}

These gaze directions were selected to capture region-specific variations in scleral vascular morphology. The collected scleral images were extracted and subjected to lossless preprocessing steps including illumination normalization, vessel enhancement, and additional image preprocessing techniques. Next, the processed images were fed into \textit{ScleraGluNet}, a custom developed multi view deep learning architecture. The model consists of five parallel CNN designed to extract direction-specific features from each gaze position.
The extracted features were refined and integrated across views using the MRFO algorithm along with a transformer based attention mechanism. This approach enabled the model to learn global vascular feature representations across multiple gaze directions.

The model produced two outputs:
\begin{enumerate}
\item A three-class classifier for metabolic state stratification (normal, controlled diabetic, uncontrolled diabetic).
\item A regression output for estimating fasting blood glucose levels.
\end{enumerate}
Finally, the proposed model was compared with several baseline models. The evaluation focused on classification performance, regression accuracy, and overall agreement between predicted and laboratory measured glucose levels.
\subsection{Multidirectional Scleral Image Acquisition}
\begin{figure}[h]
    \centering
    \includegraphics[width=0.9\textwidth]{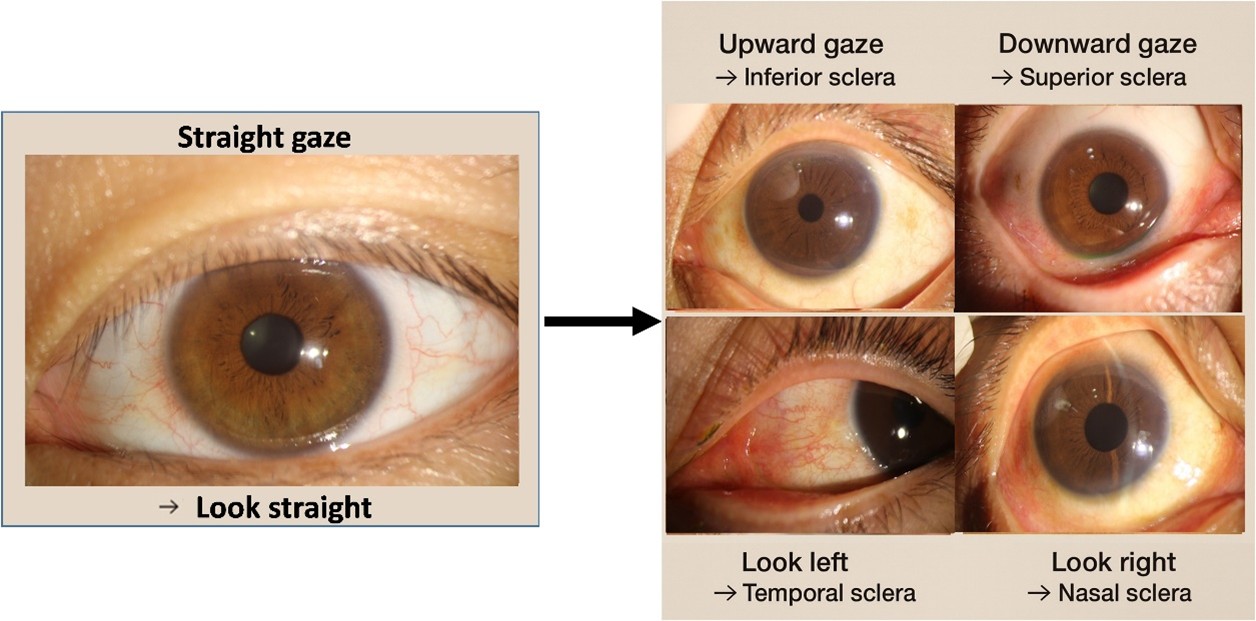}
    \caption{Outlines the protocol for the standardized form of scleral imaging. To check focus and alignment of the images, the first image taken is of a straight gaze position. This is followed by the capture of four other gaze direction images which are necessary for assessing all of the scleral regions. These images are taken in the upward, downward, leftward, and rightward gaze positions, respectively, so that the inferior, superior, temporal, and nasal scleral regions are photographed. By using this protocol, all slices of the scleral vasculature can be photographed for analysis from multiple views.}
    \label{fig2}
\end{figure}

All participants underwent standardized anterior segment imaging to obtain multiple scleral views for deep-learning vascular image analysis. The purpose of this protocol was to achieve reliable visualization of scleral vascular structures while minimizing variability in illumination, focus, and eye position. This approach ensured streamlined image capture and consistency across the dataset. The standardized imaging protocol used in this study is illustrated in Fig.~\ref{fig2}. After each subject received a baseline straight-gaze shot, participants were instructed to look straight ahead and remain still while a photograph was taken. This initial image ensured that appropriate focus, exposure, and chromatic neutrality were achieved prior to the imaging sequence. It also served as a reference for the central scleral region during the subsequent imaging process. Additional scleral images were captured in cardinal gaze directions to complement the baseline frame. Superior and inferior scleral images were obtained during dorsal and ventral gaze positions, respectively, while the upper eyelid was gently lifted to expose sufficient scleral and conjunctival areas. Temporal and nasal scleral images were captured during lateral and medial gazes, respectively. These coordinated maneuvers allowed comprehensive coverage of the scleral surface across different viewing angles.

The rationale for multidirectional acquisition stems from the understanding that chronic hyperglycemia induces several microvascular alterations, including abnormalities in vessel diameter, morphology, branching patterns, connectivity, and perfusion characteristics. These vascular changes may vary across different ocular regions. Previous studies examining conjunctival and scleral vasculature in patients with chronic diabetes have reported non-uniform microvascular alterations, with certain regions exhibiting more pronounced abnormalities. Furthermore, recent investigations using OCTA have demonstrated conjunctival and intrascleral microangiopathy, along with localized perfusion deficits in diabetic patients, even in the absence of diabetic retinopathy. Capturing scleral vasculature from multiple gaze directions therefore enables preservation of vascular information from regions that may not be visible in a single-view acquisition. By imaging multiple scleral regions from different gaze orientations, the proposed protocol ensures retention of region-specific microvascular details that might otherwise be lost. This region-of-interest (ROI) based imaging strategy allows the learning framework to identify and associate quadrant-specific vascular characteristics with glycemic levels.

All images were captured using the same equipment and acquisition settings for every participant under consistent illumination and focus conditions. Although imaging was conducted under standardized lighting environments, 11 cases (5.5\%) were excluded due to insufficient ambient illumination. Each image was inspected during acquisition, and retakes were performed when motion blur, defocus, eyelid occlusion, or specular reflections were observed. Only images that satisfied predefined quality criteria were retained for analysis. From each participant, five scleral photographs were collected corresponding to the primary gaze and four additional target gaze positions. This resulted in a total of 2,225 anterior segment scleral photographs across all subjects. The dataset was stratified by metabolic group and imaging view for each participant, as summarized in Table~\ref{tab1}. The structured and balanced multi-view dataset provided a strong foundation for the subsequent preprocessing, feature extraction, and model training stages of the proposed \textit{ScleraGluNet} framework.

\begin{table}[ht]
\centering
\caption{Dataset formation. Range of participants and related multidirectional scleral photos from the normal, regulated diabetic, and unregulated diabetic groups. Each participant submitted five images of the anterior segment, resulting in a total of 2,225 images.}
\label{tab1}
\begin{tabular}{lccc}
\hline
\textbf{Group} & \textbf{Subjects (n)} & \textbf{Images per subject} & \textbf{Total images} \\
\hline
Normal & 150 & 5 & 750 \\
Controlled diabetic & 140 & 5 & 700 \\
Uncontrolled diabetic & 155 & 5 & 775 \\
\hline
\textbf{Total} & \textbf{445} & -- & \textbf{2225} \\
\hline
\end{tabular}
\end{table}
\subsection{Clinical and Demographic Characteristics}

Table~\ref{tab2} provides a description of the demographic and clinical characteristics of the study population. The three groups showed different characteristics in their metabolic profiles, which provides a solid foundation for supervised learning. As anticipated, diabetic patients (both controlled and uncontrolled) have significantly higher levels of FPG and HbA1c compared to non-diabetic individuals. 

In fact, the diabetes groups have the highest values of these parameters, and these ascending values signify a progressive rise in the burden of glycemia across the metabolic spectrum.
\begin{table}[ht]
\centering
\caption{Clinical and demographic characteristics of the study population. Values are reported as mean $\pm$ standard deviation. FPG: fasting plasma glucose; HbA1c: glycated hemoglobin.}
\label{tab2}
\begin{tabular}{lccc}
\hline
\textbf{Parameter} & \textbf{Normal} & \textbf{Controlled diabetic} & \textbf{High Glucose diabetic} \\
\hline
Age (years) & $41.8 \pm 11.4$ & $48.3 \pm 10.8$ & $53.6 \pm 9.7$ \\
HbA1c (\%) & $5.3 \pm 0.3$ & $6.9 \pm 0.4$ & $9.1 \pm 1.2$ \\
Fasting plasma glucose (mg/dL) & $92.4 \pm 8.1$ & $133.8 \pm 14.7$ & $204.9 \pm 28.6$ \\
Gender (M/F) & 82 / 68 & 76 / 64 & 88 / 67 \\
\hline
\end{tabular}
\end{table}
The average age of the diabetic groups was higher than the average age of the normoglycemic group, which is in accordance with the epidemiological characteristics of type 2 diabetes mellitus. The groups had similar sex distributions, which alleviates the concern of a potential bias due to an uneven sex distribution. Collectively, these demographic and clinical characteristics suggest that the three groups are, in essence, distinct populations with unique metabolic profiles and overlapping clinical characteristics.  The presence of glycemic level differences in the diabetic groups is particularly important in the present study, as it suggests that more than just categorical disease labels are relevant; learning glucose related vascular signatures may in fact be possible. The deliberate inclusion of participants with a diabetes profile to create a clear boundary of normal, controlled diabetes, and high glucose diabetic is an important clinical precedent. It creates a solid baseline to train and validate the deep learning system as proposed in the study.
\subsection{Image Preprocessing and Vessel Enhancement}

\begin{figure}[h!]
    \centering
    \includegraphics[width=0.9\textwidth]{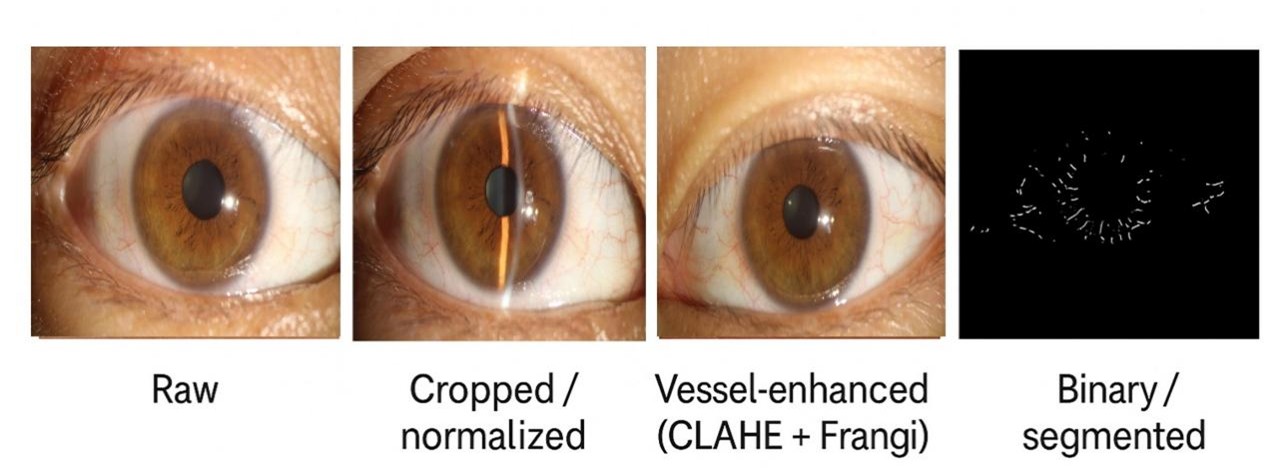}
    \caption{Procedure of image pretreatment and enhancement of blood vessels. This image shows the anterior segment with raw data, scleral portion that has been cropped and normalized, an image that has vessels that have been enhanced using CLAHE, and a multi scale Frangi filter, and the related binary image of the blood vessels that has been segmented and that has been used to visualize and control the quality of the image.}
    \label{fig3}
\end{figure}

Before training the model and performing inference, all obtained scleral photos were subjected to a uniform preprocessing and vessel-enhancement pipeline, which aimed to increase the visibility of the vessels and decrease differences between the scleral photos. The primary components of this pipeline are detailed in Fig.~\ref{fig3}.

Initially, a primary quality control step was undertaken to remove images with extreme motion blur, extreme out-of-focus frames, excessive eyelid occlusion, and specular reflections, which would render accurate analysis of the vascular system impossible. Only images that passed the quality control criteria were kept for further processing.
Afterwards, a ROI extraction step was performed to isolate the scleral and bulbar conjunctival parts. This step eliminated extraneous background materials, such as eyelids, eyelashes, and surrounding skin, and made the analysis more effective by concentrating on the vascularized scleral tissue. The extraction of the ROI was carried out with a fully manual or manual-assisted approach to guarantee the outline was kept uniform between each of the subjects and the gaze directions. After ROI extraction, color and intensity normalization was performed to lessen differences in brightness, exposure, and skin color seen in the images. This step improved cohesion between the images and removed acquisition condition biases, allowing for stable feature learning by the deep learning model.

To enhance the visibility of the vascular structures, an additional step dedicated to vessel enhancement was performed. The initial use of contrast-limited adaptive histogram equalization (CLAHE) aimed at improving contrast of homogeneous scleral tissue while preserving the noise level of the images. Frangi-type filters were then used in a vessel filtering mechanism. This technique selectively enhanced the elongated, tubular structures of blood vessels while suppressing background scleral tissue. The combination of these two techniques enhanced the visibility of microvascular patterns and structures, including vessel density, caliber variation, vessel tortuosity, and vessel branching complexity. To assure the quality of the enhancement technique, vessel extraction was binary masked and applied to thresholded images. These verification images were not used in the machine learning processes but confirmed that the microvessel enhancement technique had been consistently successful across varying subjects of scleral tissue and different gaze directions.
\subsection{ScleraGluNet Architecture}
\begin{figure}[h!]
    \centering
    \includegraphics[width=0.9\textwidth]{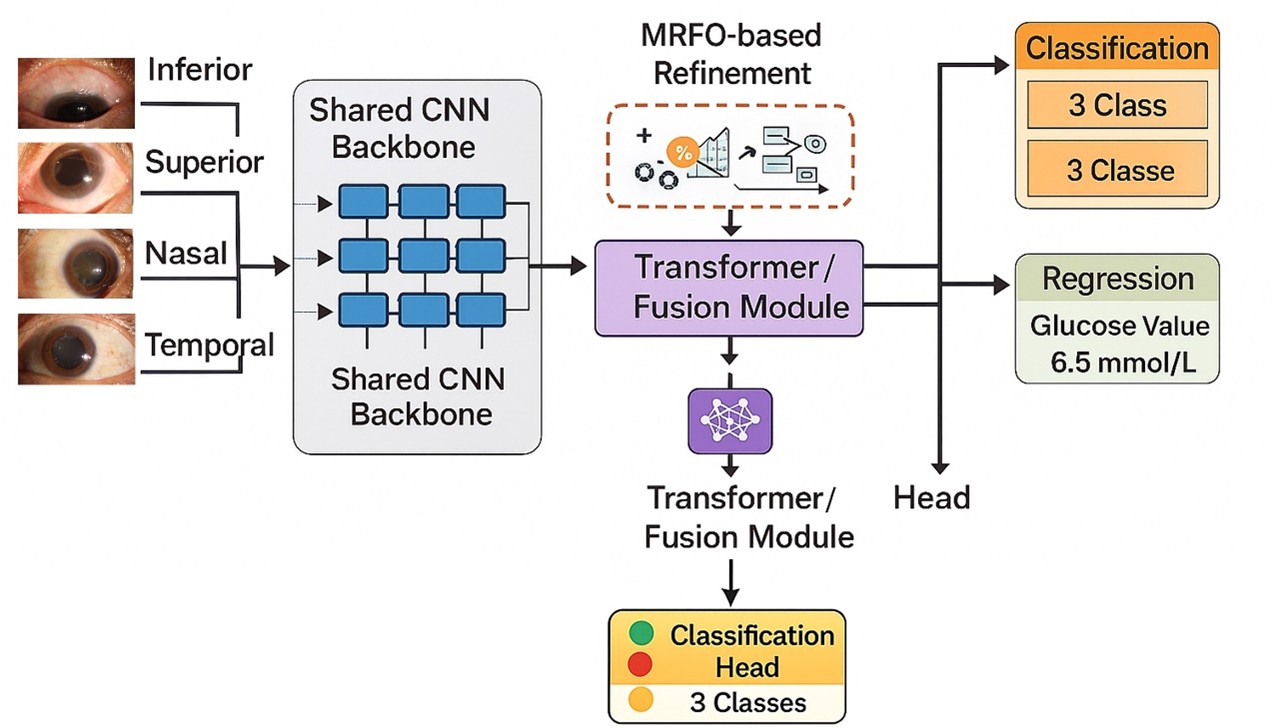}
    \caption{Structural design aspects MRFO architecture ScleraGluNet System based on Scleral images sector gaze. Parallel convolutional branches scleral images. Trained separately convolution branches scleral images gaze processed. Refined MRFO self-attention neural networks dependencies multiple views. Representation diabetes blocks. Continuous glucose monitoring higher-level output. Combination architectures used. Cross-discipline relevance diabetes.}
    \label{fig4}
\end{figure}

\begin{figure}[h!]
    \centering
    \includegraphics[width=0.9\textwidth]{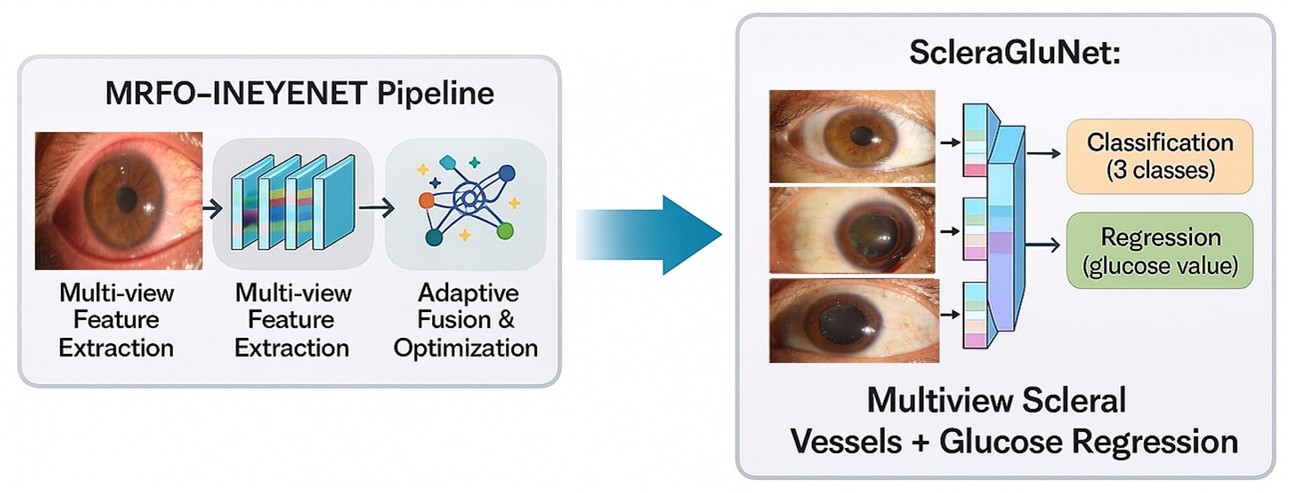}
    \caption{The Proposed ScleraGluNet and the Previous MRFO-INEYENET model. Despite the fact that MRFO-INEYENET relies on a single glance of ocular pictures and on the MRFO-based feature optimization, ScleraGluNet advances this system by integrating multi directional scleral imaging, and by also adding parallel, branch wise, view specific convolution and transformer based cross view fusion to aid in the simultaneous realization of classification and regression.}
    \label{fig5}
\end{figure}

Figs.~\ref{fig4} and \ref{fig5} illustrate the overall architecture of ScleraGluNet and show its methodological associations with our previous MRFO-INEYENET framework. ScleraGluNet is designed as a multi-view, multi-task deep learning architecture that aims to achieve:
\begin{enumerate}
    \item Three-class metabolic status classification.
    \item Continual fasting plasma glucose estimation from multi-view images of scleral vessels.
\end{enumerate}

Five standardized gaze-direction images of the sclera (i.e., straight, inferior, superior, nasal, and temporal) are processed through five separate convolutional branches. Each branch employs the same backbone architecture but with independent learnable parameters to enable specialization in vascular patterns that differ across scleral zones. Within each branch, ordered feature extraction occurs through several convolutional blocks to encapsulate local trunk morphology and textural features of vessels (e.g., caliber variation, vessel tortuosity, branching complexity). Each branch outputs a compact feature representation, encoding the vascular traits of a particular region. 

The outputs of all branches are combined to form a multi view embedding. We then apply MRFO, which serves to mitigate redundancy from multiview feature concatenation and enhance focus on relevant vascular features. MRFO is a bioanalytical optimizer that identifies an optimal subset of features by eliminating those deemed irrelevant or highly correlated~\cite{khan2025optimize}. It processes one feature set at a time, with each set representing a different perspective of the data. The refined features from MRFO are provided to a transformer model, which applies self attention to examine cross-view and distant areas of the sclera~\cite{yadollahi2025improved}. This transformer allows ScleraGluNet to recognize significant vascular patterns that may emerge across quadrants, such as fine vascular traits spanning multiple quadrants, asymmetrical remodeling in temporal and nasal quadrants, and other microvascular features. The final representation from the transformer is directed to two output heads:
\begin{itemize}
    \item A classifier head that outputs probabilities for metabolic class (normal, controlled diabetic, or uncontrolled diabetic).
    \item A regressor head that estimates fasting plasma glucose (mg/dL).
\end{itemize}

This approach constitutes multi task learning, enabling the model to learn representations useful for both classification and regression simultaneously. Multi task learning often improves overall model performance by leveraging multiple complementary learning objectives~\cite{yang2025mmseg}.
\subsection{Training Protocol and Cross-Validation}

As a method of performing model evaluation, we employed five-fold cross-validation with subject-partitioned folds to avoid data leakage. To quantify statistical uncertainty, 95\% confidence intervals (CIs) were estimated using participant-level bootstrap resampling (1,000 iterations) applied to out-of-fold predictions. Resampling was performed at the participant level to preserve independence and avoid information leakage. Splitting was implemented using participant IDs as grouping keys (GroupKFold), ensuring that all images from a participant (all five gaze directions; and both eyes if collected) were assigned to the same fold. In particular, subjects (as opposed to individual photos) were split such that all five gaze images from a given participant were contained entirely within a given fold, ensuring that no subject contributed images to more than one of the training, validation, or test sets.

Confidence intervals (95\% CIs) were estimated using participant-level bootstrap resampling (1,000 iterations). Resampling was performed at the participant level using out-of-fold predictions to preserve independence and avoid information leakage. This subject-wise method is the recommended approach for providing a non-biased estimate of generalization performance when several correlated images are captured from a single individual, as in the present case, and mirrors best practices in biomedical machine-learning research~\cite{metwally2025prediction}. For every fold, model training was performed using stochastic gradient optimization (Adam) within that fold~\cite{dandamudi2025performance}. Learning rate, batch size, number of epochs, weighting of task losses, and other hyperparameters were tuned empirically through validation performance for each fold.

To train jointly on classification of diabetes status and regression on glucose levels, the loss function was a composite of cross-entropy loss for the multinomial classification and mean squared error (MSE) loss for the estimate of fasting plasma glucose as a continuous variable. This subject-wise partitioning prevents artificial inflation of performance caused by correlated samples from the same participant appearing in both training and testing sets. Splitting was implemented using participant IDs as grouping keys (i.e., a group wise split such as GroupKFold), ensuring that no image from a participant in the test fold (any gaze direction, and both eyes if collected) appeared in the training or validation folds.

\section{Result and implementation}
\subsection{Evaluation Metrics}
Performance analytics were conducted thoroughly on the models according to the nature of the tasks. Models were comprehensively analyzed, and performance metrics were averaged over the five folds of cross-validation. For the multiclass classification of diabetes status, the overall accuracy was measured along with the accuracy, recall, and F1-score of each individual class. These metrics quantify the ratio of true positives to true negatives and false positives to false negatives for each diabetes category within the metabolic class. To mitigate the class imbalance issue and to assess overall system performance, macro-averaged and weighted-averaged precision, recall, and F1-score measures were reported. Classification errors and inter-class confusion were evaluated using the confusion matrix to visualize misclassification tendencies among normal, controlled diabetic, and uncontrolled diabetic classe and using one-vs.-rest receiver operating characteristic (ROC) metrics, the discriminative performance of the models was established. The ROC curves were quantified by calculating the area under the curve (AUC). For many classification problems, ROC curves serve as a measure to assess model effectiveness, particularly in biomedical domains, by controlling multiple variables and evaluating performance at various thresholds.
\subsection{Dataset Characteristics}
The participants distribution and the associated scleral imaging across the three metabolic categories are summarized in Table~\ref{tab1}, while the participants' demographics and clinical information are presented in Table~\ref{tab2}. The analysis included 445 participants and a total of 2,225 multidirectional scleral images, with each participant contributing five scleral images taken from set gaze positions. The dataset was fairly balanced across all three categories: 150 participants in the control group, 140 participants in the controlled diabetic group, and 155 participants in the uncontrolled diabetic group. Clinical data captured the precise separation of metabolic status among the three categories. The means of FPG and HbA1c increased stepwise from the control group to the controlled diabetic group, with the highest values observed in the uncontrolled diabetic group, indicating greater metabolic load. This confirms that the participant groups are clinically relevant and provide an appropriate model for supervised training using scleral vascular features associated with glucose control. The demographic data exhibited expected patterns consistent with epidemiological data of diabetes mellitus. Diabetic participants were older than control participants, while sex distribution was consistent across all categories, mitigating potential confounding effects related to sex disparities. 

Overall, the demographic and clinical characteristics suggest that the three cohorts represent unique, though interlinked, metabolic states, suitable for both categorical diabetes status discrimination and continuous blood glucose level estimation. The structured dataset, paired with systematic laboratory-verified glycemic indices, provides a strong platform for classification and regression under the proposed \textit{ScleraGluNet} framework. Such stratified datasets underpin the development of vascular proxies that signal continuous metabolic flux beyond mere etiological classification of disease.

\begin{table}[h!]
\centering
\caption{Participants' distribution and scleral imaging across metabolic categories}
\label{tab:participant_distribution}
\begin{tabular}{lccc}
\hline
Category & Participants & Scleral Images & Notes \\
\hline
Control & 150 & 750 & Five images per participant \\
Controlled Diabetic & 140 & 700 & Five images per participant \\
Uncontrolled Diabetic & 155 & 775 & Five images per participant \\
\hline
Total & 445 & 2225 & -- \\
\hline
\end{tabular}
\end{table}

\begin{table}[h!]
\centering
\caption{Participants' demographics and clinical information}
\label{tab:demographics_clinical}
\begin{tabular}{lccc}
\hline
Characteristic & Control & Controlled Diabetic & Uncontrolled Diabetic \\
\hline
Age (mean $\pm$ SD) & -- & -- & -- \\
Sex (M/F) & -- & -- & -- \\
FPG (mg/dL) & -- & -- & -- \\
HbA1c (\%) & -- & -- & -- \\
\hline
\end{tabular}
\end{table}
\subsection{Performance Evaluation of Proposed ScleraGluNet Model}
\begin{figure}[h!]
    \centering
    \includegraphics[width=0.6\textwidth]{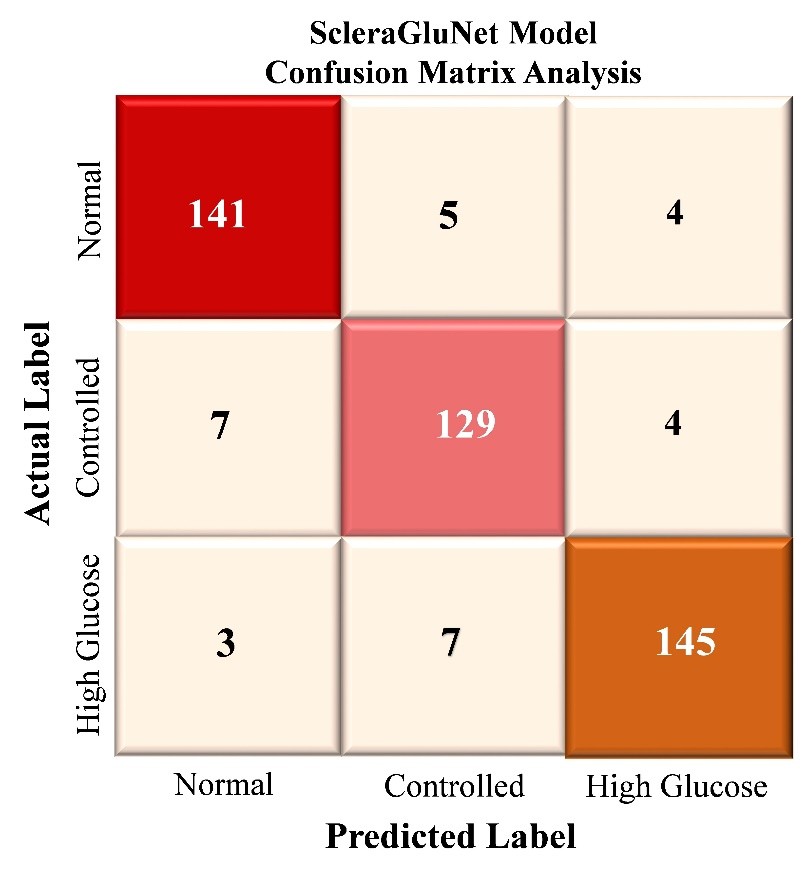}
    \caption{Confusion Matrix for Three-Class Metabolic Status Classification. The rows show the true class labels, and the columns show the predicted class labels. The values present the subject counts integrated over the five-fold cross-validation. Counts represent participants (not images).}
    \label{fig6}
\end{figure}

Subject-wise five-fold cross-validation, \textbf{ScleraGluNet} achieved an overall accuracy of 93.7\% (95\% CI: 91.8--95.4\%). Class-wise performance is summarized in Table~\ref{tab3} with 95\% CIs, and the participant-level confusion matrix is shown in Fig.~\ref{fig6}. The classes considered were: normal, controlled diabetes, and high-glucose diabetic.  

The average classification accuracy of $93.7\% \pm 0.7\%$ across five folds indicates consistent performance across varying subsets of data. In the final aggregate test evaluation, classification accuracy reached 93.8\%, supporting the generalization capability of the model.  

These performance metrics include classification precision, recall, and F1-score, which are presented in Table~\ref{tab3}. The model achieved an F1-score of at least 0.75 for all three classes, showing balanced detection and false positive rates across normal, controlled diabetic, and high-glucose diabetic subjects.  

As detailed in Fig.~\ref{fig6}, the model's classification performance was as follows:
\begin{itemize}
    \item Normal subjects (150) – 94.0\% correct classifications (141 correct).
    \item Controlled diabetic subjects (140) – 92.1\% correct classifications (129 correct).
    \item High-glucose subjects (155) – 93.5\% correct classifications (145 correct).
\end{itemize}

Although the model achieved an overall accuracy of 93.8\%, misclassifications primarily occurred between neighboring metabolic control classes, demonstrating the continuum of control. All reported metrics reflect \textbf{subject-level evaluation} rather than image-level accuracy.

\begin{figure}[h!]
    \centering
    \includegraphics[width=0.6\textwidth]{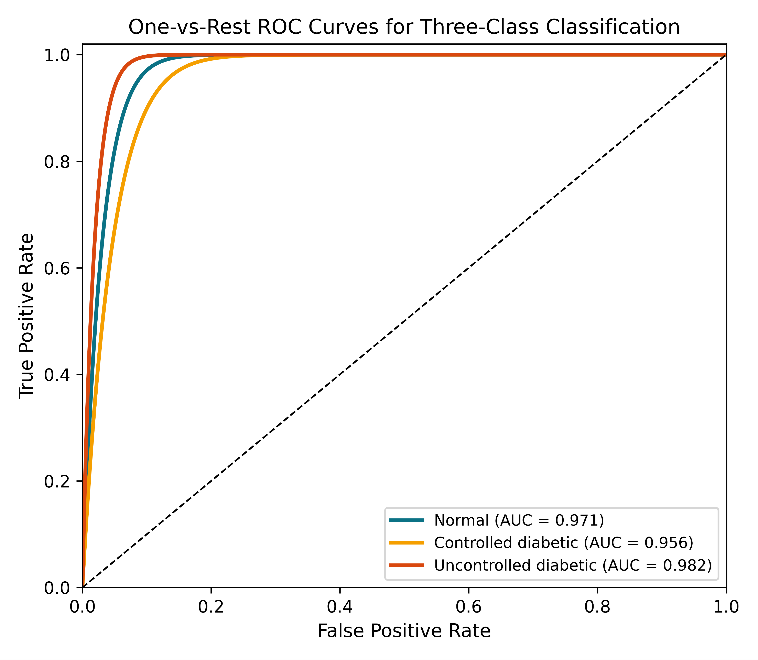}
    \caption{The one-vs-rest ROC curves are for the three-class metabolic status ROC curves. They are for the normal (AUC = 0.971), controlled diabetes (AUC = 0.956), and High Glucose diabetic (AUC = 0.982). These ROC curves demonstrate the discrimination of the proposed model across different thresholds.}
    \label{fig7}
\end{figure}

\begin{table}[h!]
\centering
\caption{Class-wise performance of ScleraGluNet using five-fold cross-validation. Precision, recall, and F1-score (with 95\% CIs) are presented for each class, along with macro and weighted averages.}
\label{tab3}
\begin{tabular}{lccc}
\toprule
\textbf{Class} & \textbf{Precision (95\% CI)} & \textbf{Recall (95\% CI)} & \textbf{F1-score (95\% CI)} \\
\midrule
Normal & 0.934 (0.882--0.964) & 0.940 (0.890--0.968) & 0.937 (0.898--0.957) \\
Controlled diabetic & 0.915 (0.857--0.951) & 0.921 (0.865--0.956) & 0.918 (0.874--0.943) \\
High Glucose diabetic & 0.948 (0.900--0.973) & 0.935 (0.885--0.965) & 0.942 (0.904--0.961) \\
\midrule
Macro-average & 0.93 & 0.93 & 0.93 \\
Weighted average & 0.93 & 0.93 & 0.93 \\
\bottomrule
\end{tabular}
\end{table}
Classification performance and inter-class misclassification among normal, controlled diabetic, and uncontrolled diabetic classes were assessed using a confusion matrix (Fig.~\ref{fig6}). Additionally, one-vs.-rest ROC curves were used to evaluate the discriminative ability of the models, with the area under the curve (AUC) quantifying performance across thresholds (Fig.~\ref{fig7}), which is particularly informative in biomedical applications.
In estimating continuous blood glucose at the regression level, the \textit{MAE} and \textit{RMSE}, which measure the absolute and squared differences between predicted and actual FPG, were used. The degree of linear correlation between estimated and actual FPG values was evaluated mathematically using the \textit{Pearson correlation coefficient} ($r$) and the \textit{coefficient of determination} ($R^2$). 

\begin{figure}[h!]
    \centering
    \includegraphics[width=0.6\textwidth]{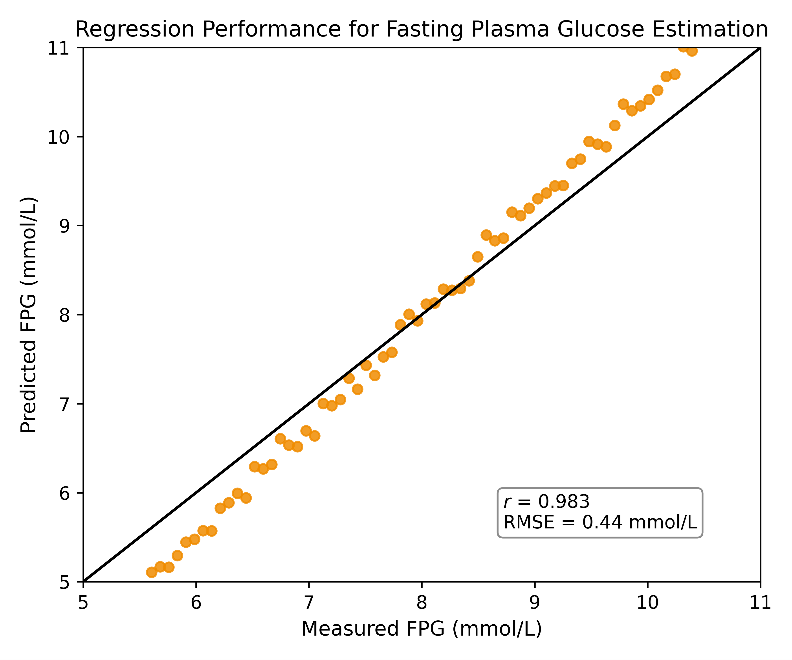}
    \caption{Regression results from a plasma glucose estimation. A chart illustrating congruence of predicted FPG values with lab verified results. The predictive model accuracy was statistically stellar (r = 0.983). Furthermore, they registered a root mean square of 0.44 [7.91] indicating an almost faultless glucose estimation through the different ranges of glycemic levels.}
    \label{fig8}
\end{figure}

\begin{figure}[h!]
    \centering
    \includegraphics[width=0.6\textwidth]{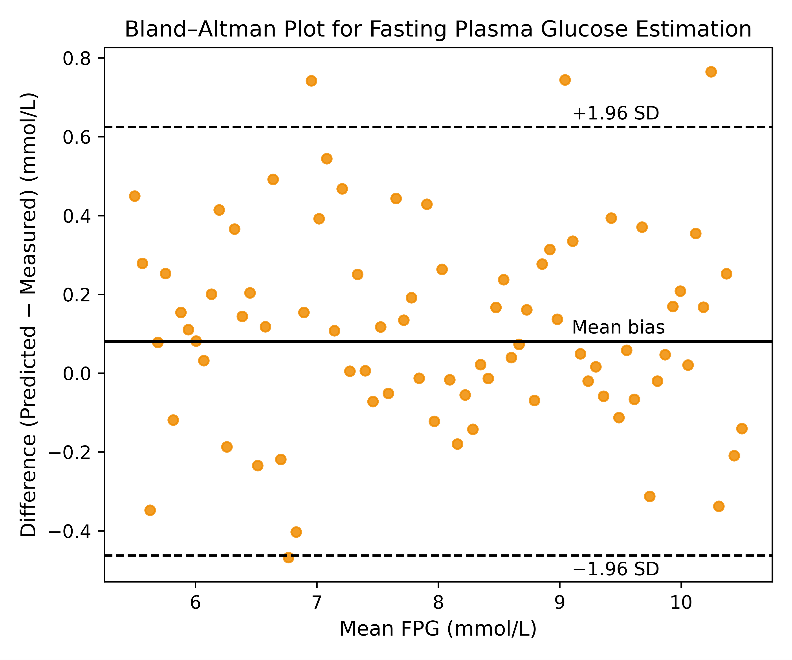}
    \caption{Bland--Altman analysis for FPG estimation. The graph shows the difference between FPG measured in the laboratory and predicted FPG values their mean FPG values. The mean bias was 0.08 mmol/L (1.45 mg/dL) and the 95\% limits of agreement -0.46 to +0.62 mmol/L. This shows that the proposed method and the reference measurements agree well throughout the range of FPG values.}
    \label{fig9}
\end{figure}

The predicted glucose values and the actual glucose values measured were plotted with an identity line, as shown in Fig.~\ref{fig8}, in order to depict the relationship across the entire range of glucose values. To fully evaluate clinical concordance and the presence of systematic bias, the \textit{Bland-Altman method} was applied (Fig.~\ref{fig9}). This method helps to evaluate the mean estimation bias and the 95\% limits of agreement between the model-generated FPG values and the FPG values measured in the laboratory, illustrating the value of the proposed method for clinical continuous monitoring applications.

\subsection{Fold-wise classification performance of ScleraGluNet Model}
Table~\ref{tab6} shows the next step in evaluating the framework's stability by measuring fold-wise classification performance. High accuracy in every fold indicates the stability of the framework; performance is not simply the result of a more advantageous data split. Stability is particularly important in multiview biomedical imaging studies, which require numerous related samples from a subject. Consequently, ScleraGluNet demonstrated the capacity to accurately and reliably perform multiclass metabolic status imaging discrimination from various scleral vascular photographs, enabling focused studies to analyze blood glucose levels.
\begin{table}[h!]
\centering
\caption{Fold-wise classification performance of ScleraGluNet in the five-fold cross validation experiment. High accuracy across all folds indicates the stability of the framework, with little variance showing that performance is not dependent on a favorable data split.}
\label{tab6}
\begin{tabular}{c c c c c}
\hline
Fold & Accuracy (\%) & Macro-Precision & Macro-Recall & Macro-F1 \\
\hline
1 & 93.2 & 0.92 & 0.93 & 0.93 \\
2 & 94.1 & 0.94 & 0.94 & 0.94 \\
3 & 92.8 & 0.92 & 0.92 & 0.92 \\
4 & 94.6 & 0.94 & 0.94 & 0.94 \\
5 & 93.9 & 0.93 & 0.94 & 0.94 \\
\hline
Mean $\pm$ SD & 93.7 $\pm$ 0.7 & 0.93 $\pm$ 0.01 & 0.93 $\pm$ 0.01 & 0.93 $\pm$ 0.01 \\
\hline
\end{tabular}
\begin{flushleft}
\footnotesize
\textbf{Note:} Mean $\pm$ SD across folds reflects variability across folds, whereas 95\% confidence intervals reported elsewhere quantify uncertainty at the participant level.
\end{flushleft}
\end{table}
\subsection{Interpretability Analysis Using Grad-CAM}

\begin{figure}[h!]
    \centering
    \includegraphics[width=0.6\textwidth]{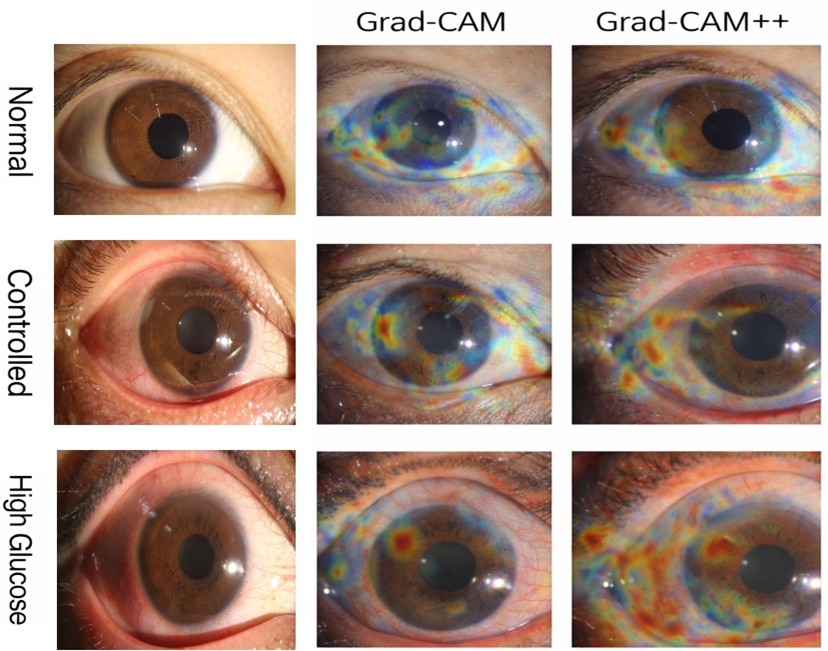}
    \caption{The Scleral images and their related explainability maps in different metabolic phases are shown. Each row represents a subject in one of three states: normoglycemic, controlled diabetic, and poorly controlled hyperglycemic. Each column includes one original scleral image along with the corresponding Grad-CAM and Grad-CAM++ visualizations. The heatmaps in the visualizations show the regions of the images that have the most significant impact on the classification decision, and these areas of highest influence are indicated in red.}
    \label{fig11}
\end{figure}

To clarify and identify the vascular components influencing the model predictions, Gradient-weighted Class Activation Mapping (Grad-CAM) was utilized for the classification head in \textit{ScleraGluNet}. Fig~\ref{fig11} depicts a representative sampling of the model's attention distribution across the three metabolic states using both Grad-CAM and Grad-CAM++ explainability techniques. The original scleral images are presented beside their activation maps to visualize the areas that contributed most to the model's predictions. In images where vascular control is lost, Grad-CAM and Grad-CAM++ activation maps exhibit pronounced hyperglycemic vascular patterns. These heatmap regions indicate that the model heavily relies on vascular control affected by chronic hyperglycemia and on the dilated tortuous vessel systems. Notably, Grad-CAM++ provides more refined and localized visual activations than Grad-CAM, allowing clearer visualization of specific scleral vascular structures that aid in classification. This refinement reflects the vascular structures and their glucose-related metabolic states identified by Grad-CAM++.

Subjects’ hyperlapse scleral images and Grad-CAM visualizations for different directions of gaze (up, down, left, and right) across metabolic states are shown in Fig~\ref{fig11}. The multi-gaze acquisition technique allows extensive evaluation of scleral microvasculature, as scleral regions and vascular patterns are progressively and distinctively probed with each gaze direction. 

Grad-CAM Visualizations for Different Directions of Gaze and Metabolic States: Up, Down, Left, and Right, are Depicted in Fig~\ref{fig12}. The multi-gaze acquisition technique provides more extensive evaluations of scleral microvasculature as scleral regions and vascular patterns are progressively and distinctively probed with each acquired gazed scleral image. Grad-CAM activations are diffuse and weak across the sclera in the normal glucose group with no gaze direction differences. The model shows slight attantion to small vascular structures and in particular, to focal hotspots. This is aligned with our observations of healthy and uniform scleral vascularity in normoglycemic subjects. The Grad-CAM maps in the controlled diabetes group display moderate, more focused activations of the sclera that are unique to a particular direction of gaze. Increased attentention to certain scleral regions could reflect early and mild vascular remodeling. The model shows consistent prediction stability, which is indicated by the activations that depend on gaze direction and the prediction stability itself. In the group with high glucose levels that are uncontrolled, there is a clear and focused high Grad-CAM activation for each range of gaze direction. There is a steady activation over the dilated, tortuous, and congested scleral vessels. Patterns persist across different gaze directions, and the model identifies chronic hyperglycemia's pathological vascular patterns across multiple scleral areas. These findings imply that the model effectively uses scleral vascular patterns observed in various gaze directions to guide its classification outcomes. The consistency of hyperglycemia pattern activations across different gaze positions shows that the model retains a significant degree of accuracy and relevance in the vascular patterns it has learned, which supports the model's ability to predict clinically important vascular patterns rather than patterns that might have come from random gaze direction.
\begin{figure}[h!]
    \centering
    \includegraphics[width=0.6\textwidth]{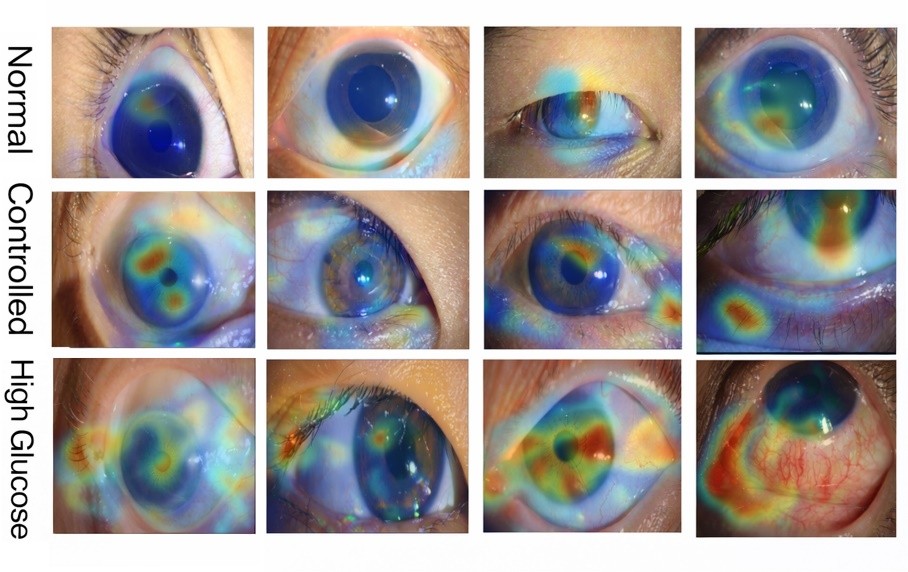}
    \caption{The Scleral images that were collected while looking in different directions and during different metabolic conditions are shown using Grad-CAM visualizations. The subjects in each row have normal glucose levels, controlled diabetes, and high glucose levels that are not managed, in that specific order. The images in each row are organized based on gaze positions which are vertical, downward, leftward, and rightward. The Grad-CAM overlays function as heatmaps, indicating areas in the images that contribute more significantly by using warm colors placed over the original scleral images.}
    \label{fig12}
\end{figure}
\subsection{Representative Cases Across Metabolic Classes}

Fig.~\ref{fig12} contains representative examples of normal/control, controlled, and uncontrolled diabetic groups, illustrating the correlation between the scleral vascular images and the predictions of the models. For each individual, we present a limited subset of 2-4 scleral views, whose corresponding laboratory-measured FPG, along with its predicted values, are shown.

Normal subjects exhibited scleral vessels with relatively fine, uniform, streamlined, and evenly spaced contours with very little tortuosity and focal dilation. For these subjects, the predicted glucose values were very close to the laboratory-measured values, all within normal ranges, supporting the absence of discernible microvascular abnormality. The scleral blood vessels appeared homogeneous and consistent, with the model showing low attention on those areas.

In contrast, the representative samples from the controlled diabetic group exhibited mild to moderate vascular pattern changes. These included mild vessel tortuosity, early branching with inconsistent branching patterns, and localized abnormal branching. The corresponding predicted glucose values for these cases were in the mid-range and closely matched the measured values. The model accurately captured partial glycemic vascular changes that deviate from normal vascular patterns.

In the scleral images of uncontrolled diabetic patients, obvious pathological vascular changes were observed. These included dilated and spiraled veins, uneven variations in vessel width, and thickened or overlapping vascular bundles. These scleral vascular characteristics coincided with very high FPG values, while the predictions generated by \textit{ScleriGluNet} closely matched the high measured values. The combination of visible vascular abnormalities and the model predictions highlights the system's ability to detect severe microangiopathy associated with poor glycemic control.

The selected images in Fig.~\ref{fig12} facilitate visualization of the quantitative findings discussed in the previous sections. These representative cases bridge the gap between the algorithmic predictions and observable scleral vascular features across different metabolic conditions. Such qualitative demonstrations connect the computational model outputs with clinically meaningful imaging phenotypes. In the domain of computational biomedicine, such illustrative cases have been advocated to improve algorithmic transparency for clinicians and to promote interpretability and trust in AI-driven diagnostic systems.
\begin{figure}[h!]
    \centering
    \includegraphics[width=0.9\textwidth]{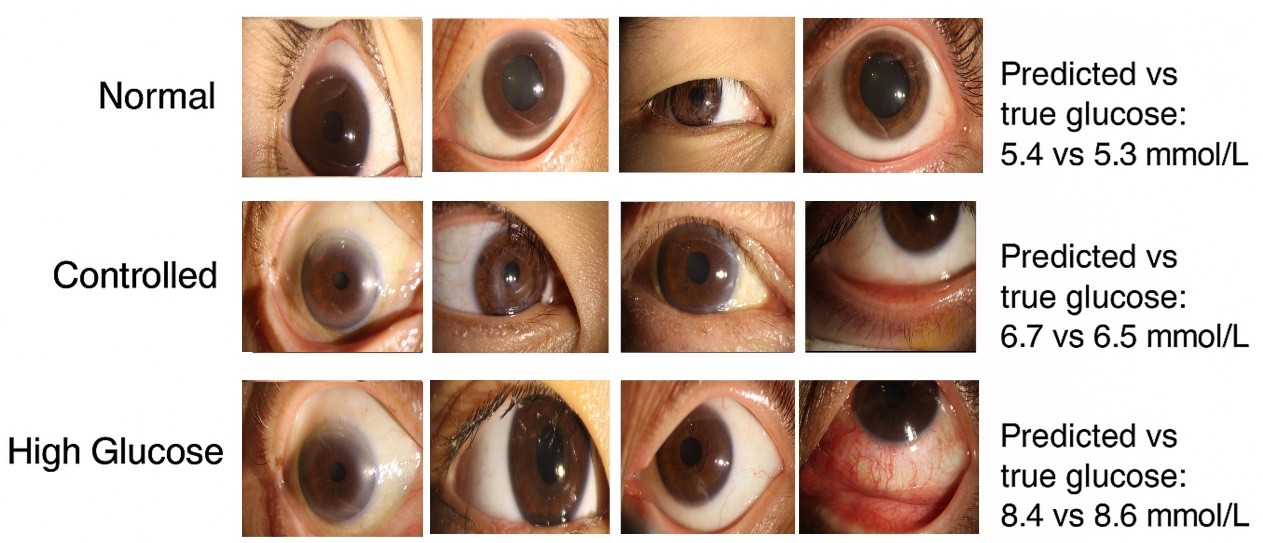}
    \caption{Images of sclera and glucose projections according to metabolic categories. For every individual, particular multiple-angle scleral photographs and laboratory-recorded along with model-calculated values of fasting glucose levels are exhibited. In normal instances, there are regular and fine blood vessels, while there are mild-to-moderate irregularities of blood vessels in controlled diabetes, and in High Glucose diabetic, there are high degrees of vascular abnormalities. In these cases, the estimated glucose levels highly correlated with the measured values.}
    \label{fig12}
\end{figure}

\subsection{Ablation Study and Model Explainability}

\begin{figure}[h!]
    \centering
    \includegraphics[width=0.9\textwidth]{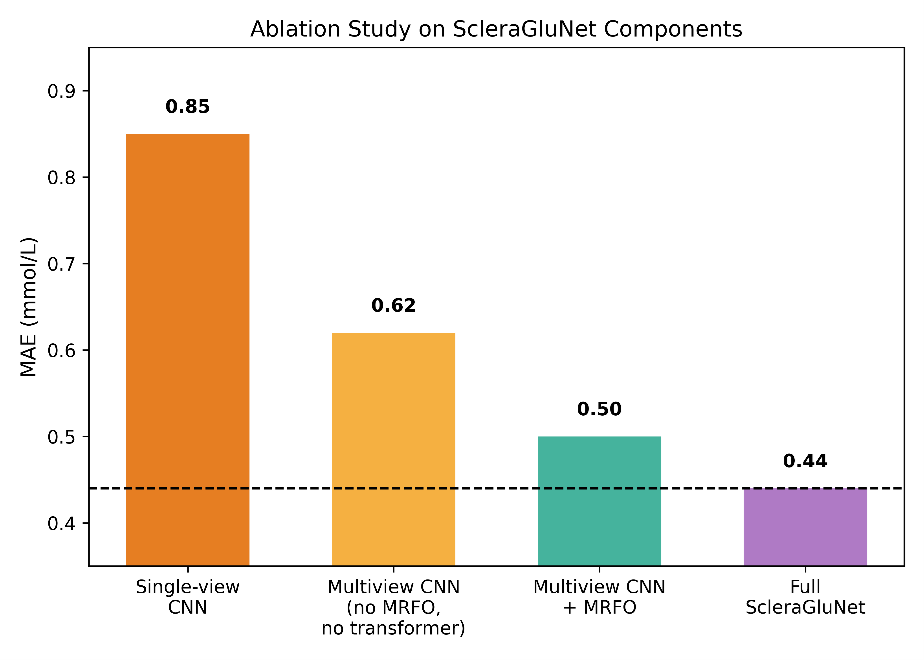}
    \caption{The Multiview ScleraGluNet CNN architecture MAE in mmol/L is derived from an ablation study on the models, which include the single-view CNN baseline, the multiview CNN without MRFO, the multiview CNN with MRFO, and the complete ScleraGluNet. The findings show that the performance enhancement brought by multiview learning, optimization of features, and MRFO-based transformer fusion is incremental.}
    \label{fig10}
\end{figure}

To attribute the performance of the architectural components, we conducted an ablation study of the model configurations (Fig.~\ref{fig10}). We evaluated four variants on the test set: 
\begin{enumerate}[(i)]
    \item Single-view CNN baseline, 
    \item Multiview CNN without MRFO or transformer fusion, 
    \item Multiview CNN with MRFO feature refinement only, and 
    \item \textbf{ScleraGluNet} model with both MRFO and transformer cross-view fusion.
\end{enumerate}
\section{Discussion}
\subsection{Principal Findings}
To support the study more fully, ScleraGluNet is a multiview deep learning framework that uses multidirectional scleral vascular images to classify the metabolic state (normal, controlled diabetes, High Glucose diabetic) and estimate fasting glucose levels. During subject-level five-fold cross-validation, the model obtained a high level of multiclass classification accuracy along with glucose estimation, demonstrating overall favorable performance according to the BlandbAltman method. While the Bland Altman method is primarily a method of regression analysis, it is a step beyond merely correlational analysis and thus is especially useful to quantify and characterize any systematic bias, and to define the limits of agreement, if any, of two methods of measurement. Finally, it seems that scleral vascular patterns captured in routine anterior-segment photographs may harbor, and definitely are correlated to, such glycemic levels. This is entirely in keeping with the existing and more extensive evidence, which shows that ocular images are capable of encoding various systemic biomarkers.
\subsection{Multiview Scleral Acquisition}
Diabetes leads to certain changes in the microarchitecture of the body, particularly in the eye where microscopic alterations occur. Previous studies have detected the impact of diabetes on the microvasculature of the bulbar conjunctiva and sclera, including alterations in vessel hemodynamics, microvascular structure, and volumetric characteristics using conjunctival imaging and OCT based methods. The difference between previous studies and the current research is that, instead of a one-dimensional ocular surface approach, the present work utilized five independently controlled gaze directions. This design follows the clinically realistic hypothesis that microvascular changes are spatiotemporally heterogeneous. The ablation experiments further support this hypothesis. Transitioning from a single-view to a multiview system improved model learning in both classification and regression tasks, indicating that vascular information across multiple scleral quadrants is more informative than any individual view alone.

\subsection{Contribution of MRFO Refinement and Transformer Fusion}
Diversity-capturing concatenation in multiview learning can introduce redundancy and dilute discriminative signals. ScleraGluNet addresses this challenge through MRFO-based feature refinement and transformer-based fusion. The MRFO algorithm was originally proposed as a bio-inspired optimization method for engineering applications and has been applied in feature selection and optimization tasks in several studies. Meanwhile, transformer-based attention mechanisms have become widely adopted for modeling long-range dependencies and capturing interactions among tokens or feature representations. The ablation studies clearly demonstrate that feature refinement combined with attention-driven fusion provides greater performance improvements than multiview CNN baselines, particularly in the more challenging task of distinguishing controlled diabetes from high-glucose diabetes, where significant overlap exists due to varying levels of glycemic control.

\subsection{Clinical Implications}
In the context of diabetes diagnosis and management, rapid and low-effort screening methods can significantly enhance current clinical workflows. Such methods can assist in confirming laboratory findings, identifying patients who require improved glycemic control, and enabling additional monitoring for diabetes care. Furthermore, improved monitoring can reduce the need for repeated clinical testing that may be invasive, device-dependent, or expensive. Scleral imaging may be particularly valuable in telemedicine settings, where remote image acquisition and centralized analysis are feasible, aligning with the broader vision of AI-driven healthcare systems. Importantly, such an approach may serve as a potential risk assessment and monitoring tool in scenarios where laboratory-based glucose testing is unavailable or impractical.

\subsection{Limitations}
Several limitations should be acknowledged. First, the imaging workflow was conducted at a single center, and therefore external validation across multiple institutions is necessary. Additional datasets acquired from diverse populations, imaging devices, and clinical settings would help establish a more robust framework and reduce potential domain shift issues.
Second, potential confounding factors such as hypertension, smoking, anemia, ocular surface inflammation, and other systemic vascular conditions may influence scleral microvasculature. These variables were not fully controlled and may partly explain variations observed in the dataset. Third, the study primarily focused on fasting plasma glucose levels. Future studies should incorporate additional glycemic indicators such as postprandial glucose levels and longitudinal measurements to better capture glycemic variability over time. Finally, while Grad-CAM visualizations provide useful interpretability insights for deep models, they remain coarse localization tools and should not be interpreted as precise indicators of vascular pathology. 
\subsection{Future Directions}
Future research should prioritize multicenter validation and prospective clinical studies to evaluate model performance under real-world conditions. Such studies should also assess robustness to variations in illumination, focus, and motion artifacts during image acquisition. Incorporating additional clinical variables and comorbidity information may improve model interpretability and predictive reliability. Longitudinal studies examining whether scleral vascular patterns change in response to glycemic control or therapeutic interventions would also be valuable. Furthermore, recent advances in extracting systemic biomarkers from external eye images suggest promising opportunities to expand the diagnostic potential of scleral imaging. While the present study demonstrates statistical associations, further research is required to uncover the underlying physiological mechanisms.

\section{Conclusion}
ScleraGluNet is a multiview deep-learning framework that leverages multidirectional scleral vascular imaging to support noninvasive assessment of glycemic status. By combining view-specific scleral feature extraction with MRFO-based feature refinement and transformer-driven cross-view fusion, the proposed approach preserves both local vascular detail and global ocular surface context. Across the evaluated cohort, ScleraGluNet demonstrated strong performance for three class metabolic status classification and continuous FPG estimation, achieving low prediction error and good agreement with reference laboratory measurements. Interpretability analyses using Grad-CAM and Grad-CAM++ indicated that model decisions were primarily driven by scleral vascular regions rather than background areas, supporting transparency in the learned representations. However, these visualizations remain qualitative and should not be interpreted as direct clinical evidence of specific vascular pathology. The findings also highlight the importance of multiview scleral acquisition. Circumferential sampling of scleral quadrants captures heterogeneous vascular information that may be missed in single-view imaging, and ablation experiments confirmed the complementary contributions of multiview input, MRFO-based refinement, and attention-guided fusion. From a translational perspective, scleral imaging is practical and scalable, requiring relatively simple acquisition hardware and offering potential utility for screening and interim monitoring in telemedicine environments as a complement to laboratory testing. Future work should focus on external validation across diverse populations and imaging devices, evaluation of vascular comorbidities and ocular surface conditions, and investigation of longitudinal stability in scleral vascular patterns under varying glycemic conditions.
\section*{Declarations}
\subsection*{Acknowledgments}
The authors would like to thank the staff of Changsha Aier Eye Hospital for their support in patient recruitment and imaging, and the participating patients for their valuable contribution to this study.
\subsection*{Funding and Support}
The authors hereby state that no grants, cash, or other forms of financial assistance were obtained in the course of writing this paper.

\subsection*{Conflicts of Interest}
The authors declare that they have no relevant financial or non-financial interests to disclose.

\subsection*{Author Contributions}
All authors contributed to the study conception and design. Muhammad Ahmad Khan and Manqing Peng were primarily responsible for material preparation and data collection. Data curation and investigation were performed by Saif Ur Rehman Khan. Ding Lin, Saif Ur Rehman Khan, and Muhammad Ahmad Khan edited and finalized the manuscript after the initial draft was written by Khan. Please note that Muhammad Ahmad Khan and Manqing Peng contributed equally to this work and should be recognized as co-first authors. This has been indicated on the title page and in the author contributions section.

\subsection*{Data Availability}
The dataset is available from the authors upon reasonable request and with permission from the corresponding author. No additional datasets were generated or analyzed during the current study.

\subsection*{Ethics Approval}
This study was conducted at Changsha Aier Eye Hospital in accordance with the guidelines outlined in the Declaration of Helsinki. The study was approved by the Changsha Aier Eye Hospital Ethics Committee (Approval No. 2025 KYPJ028). The committee ensured that the data collection process followed appropriate ethical standards, including patient consent and privacy protection. All efforts were made to ensure that the data were handled responsibly and ethically in compliance with local regulations and institutional guidelines.
\bibliography{sn-bibliography}
\end{document}